\documentstyle[epsf,seceq,twoside]{ptptex}
\notypesetlogo  
\title{Possible Evidence for a Chiral Particle $B_0^\chi$,\\
Scalar Meson in the $B$ Meson System}
\author{%
Daiki {\sc Ito}, Muneyuki {\sc Ishida}$^*$, Shin {\sc Ishida},
Toshihiko {\sc Komada}$^{**}$, Tomohito {\sc Maeda} and Ichiro {\sc Yamauchi}$^{***}$ 
}
\inst{%
Research Institute of Quantum Science, 
College of Science and Technology\\
Nihon University, Tokyo 101-0062, Japan\\
$^*$Department of Physics, Tokyo Institute of Technology, 
Tokyo 152-8551, Japan\\
$^{**}$Department of Engineering Science, Junior College Funabashi Campus \\ 
  Nihon University, Funabashi 274-8501, Japan\\
$^{***}$ Tokyo Metropolitan College of Technology, 
Tokyo 140-0011, Japan
}
\abst{%
In the covariant level-classification scheme of hadrons proposed recently, 
the existence of ``chiral particles", scalar and axial-vector mesons 
as the partners of the ground-state pseudo-scalar and vector mesons, respectively,
were predicted in the $D$ meson and $B$ meson systems, 
realizing a linear representation of chiral symmetry. 
In this work we reanalyze the $B^0\pi$ mass spectra through $Z^0$ decays 
to show a possibility for the scalar $B_0^\chi$.
}

\begin{document}
\maketitle

\setcounter{tocdepth}{4}

\section{Introduction}

The level classification of hadrons has been done rather successfully on the basis of the 
non-relativistic quark model(NRQM) for these 4 decades. However, this non-relativistic 
scheme seems now to be confronted with a serious difficulty: Recently existence\cite{sca,HK} of the light-mass
iso-scalar scalar meson to be identified with $\sigma$\cite{rfsigma,rf555}, 
the chiral partner of the $\pi$ meson as a 
Nambu-Goldstone boson, has been accepted widely\cite{2000}. 
Successively possible existence of the $\kappa$(900)
meson\cite{kappa}, iso-spinor scalar meson, has been pointed out. Then we are able to identify naturally the $\sigma$-meson
nonet\cite{rfmw} with the members $\{ \sigma (600),\kappa (900),a_0(980),f_0(980)  \}$.

The mass values $(\le$ 1GeV) of this scalar nonet are in the region of $q\bar q$ ground states in NRQM,
while there are no seats for scalars in this model. This has aroused recently a hot controversy on what the
quark configuration of $\sigma$ nonet is.

On the other hand we have proposed a new level-classification scheme\cite{rfCLC,u12}, which has a relativistically covariant
framework and in conformity with the approximate chiral symmetry. 
In this new classification scheme the ``chiral states", which are out of the framework of
the NRQM, are expected to exist in the lower mass region, and the $\sigma$-nonet
is naturally assigned as the $(q\bar q)$ relativistic $S$-wave chiral state. In the heavy-light quark meson system
it is expected that the approximate chiral symmetry on the light quark is valid, and the new level-classification scheme
anticipates existence of the chiral states\cite{hl}, scalar mesons and axial-vector mesons, in addition to the conventional states,
pseudo-scalar and vector mesons, in the ground states. 

In the previous work\cite{D1chi} we have shown a possible evidence for existence of a chiral particle $D_1^\chi$,
axial-vector in the $D$ meson system. The purpose of this paper is to investigate the possibility for existence
of another chiral particle $B_0^\chi$, scalar in the $B$ meson system, by reanalyzing some experimental data
reported previously.
 
\section{Experimental Data and Project for Reanalysis}

\hspace*{-0.8cm} ({\it Experimental data to be analyzed})

As the experimental data to be reanalyzed we choose the mass spectra 
of $B\pi$ system obtained through $Z^0$ boson decay
\begin{eqnarray}
 e^++e^- &\rightarrow& Z^0 \rightarrow B\pi + \cdots ,\ \ \label{eq1}
\end{eqnarray}
by L3 collaboration\cite{L3} with some statistical accuracy.
We also apply the data by ALEPH collaboration\cite{ALEPH} with low statistical accuracy
for suplementary analysis. The original data are inclusive and the relevant exclusive mass spectra
of $B\pi$ system are obtained by subtracting from the original ones the backgrounds of the form
\begin{eqnarray}
B.~G.~ &=& P_1 (\Delta M)^{P_3}\ {\rm exp}\left[ P_4(\Delta M) + P_5(\Delta M)^2 + P_6(\Delta M)^3   \right]
\ \  \label{eq2}\\
\Delta M &\equiv& M_{B\pi}-P_2,\ \ \nonumber
\end{eqnarray}
with fitting parameters $P_i\ (i=1,2,\cdots ,6)$.
We shall apply the same formula to Eq.~(\ref{eq2}) for the backgrounds,
taking into account the possible effects of intermediate $B_0^\chi$ production.

\hspace*{-0.8cm} ({\it Relevant intermediate resonances})

The conventional two $(b\bar q)-P$ wave mesons may contribute directly, as intermediate states, 
to the final $B\pi$ system with the respective angular momenta $l=0,2$ as
\begin{eqnarray} 
{\rm Direct\ Reson.\ Process}\ \ \ \ \  B_0^* & \rightarrow & B + \pi\ \ (S-{\rm wave})\ ,\ \  \label{eq3} \\
                    B_2^* & \rightarrow & B + \pi\ \ (D-{\rm wave})\ ,\ \  \label{eq4} 
\end{eqnarray}
where $B_0^*$ and $B_2^*$ have, respectively, ${}^{j_q}L_J={}^{1/2}P_0$ and ${}^{3/2}P_2$
$( {\mib j}_q={\mib S}_q+{\mib L}$ is the total amgular momentum of the light quark).

In this work, a possible direct contribution from, in addition to the above conventional resonances, 
the chiral scalar meson $B_0^\chi$ is taken into account as 
\begin{eqnarray}
{\rm Direct\ Reson.\ Process}\ \ \ \ \   B_0^\chi & \rightarrow & B + \pi \ (S-{\rm wave})\ , \ \label{eq5} 
\end{eqnarray}
Concerning a possible candidate for the chiral particle $B_0^\chi$,
some excess structure over the backgrounds around the mass $m\sim 5550$ MeV,
which is common to both the data of L3 and ALEPH,
is identified as being due to production of $B_0^\chi$.

In the relevant experiment the low energy $\gamma$ was unable to be observed.
Accordingly we must take into account the background process from the
intermediate resonances, decaying into $B^*+\pi$ (successively $B^*$ decays into $B$ and missing $\gamma$). 
\begin{eqnarray} 
{\rm Backgd.\ Reson.\ Process}\ \ \ \ \   B_1^* & \rightarrow & B^* + \pi\ \ (S-{\rm wave})\ ,\ \  \label{eq6} \\
                   B_1 & \rightarrow & B^* + \pi\ \ (D-{\rm wave})\ ,\ \  \label{eq7} \\  
                   B_2^* & \rightarrow & B^* + \pi\ \ (D-{\rm wave})\ ,\ \  \label{eq8} \\ 
                  & & B^* \rightarrow B + \gamma\  .\ \ \label{eq9}
\end{eqnarray}

\hspace*{-0.8cm} ({\it Formulas for analysis})

We shall apply the VMW method in our relevant case,
where the absolute amplitude squared is given by
\begin{eqnarray}
|M(s)|^2 &=& \{ | r_1\Delta_{B_0^\chi}(s)+r_2e^{i\theta}\Delta_{B_0^*}(s)|^2 
               +| r_3\Delta_{B_2^*}(s)|^2   \}   \nonumber\\
               & & + \{ | r_4\Delta_{B_1^*}(s)|^2 + | r_5 \Delta_{B_2^*}(s)|^2 
                 +| r_6\Delta_{B_1}(s)|^2   \}   \label{eq10} \\
  \Delta_i(s) & \equiv & \frac{-m_i \Gamma_i }{s-m_i^2+im_i\Gamma_i}
\ \ \ \ \ \ (i\ {\rm denoting\ respective\ resonances})\ . \label{eq101}
\end{eqnarray}
Here the first (second) term represents the contributions from the direct (background) resonance
processes Eqs.~(\ref{eq3}-\ref{eq5}) (Eqs.~(\ref{eq6}-\ref{eq8}) ), the $m_i(\Gamma_i)$
are the mass(width) of relevant resonances, and $r_i$ represent their production strength.
In Eq.~(\ref{eq10}) a possible interference effect between the two direct decay processes
of the $B_0^*$ (Eq.~(\ref{eq3})) and of the $B_0^\chi$ (Eq.~(\ref{eq5})) are taken into account,
while no interference effects among any background process Eqs.~(\ref{eq6}-\ref{eq8})
are expected.

\hspace*{-0.8cm} ({\it Project for applying $\chi^2$-analysis})

We shall perform the two kinds of $\chi^2$-fitting of the mass spectra, the one with $B_0^\chi$
and the other without $B_0^\chi$, and compare with their results.
Our main interest for reanalysis in this work is to search for a possibility of existence
of $B_0^\chi$. Accordingly we shall follow to the projects of original analysis,
modifying them only in the places to be related with this existence, and our projects are:\\
i) Backgrounds; \ \ \ \ in the case without $B_0^\chi$ their parameters, $P_1$ through $P_6$,
are determined so as to reproduce the original background curve given by L3, while 
in the case with $B_0^\chi$ the original curve is revised in the region close to threshold
where the possible effects of $B_0^\chi$ may be included.\\
ii) mass and width of resonances; \ \ \ \ their values of $m_i$ and $\Gamma_i$ appeared in 
Formulas (\ref{eq10}) are restricted in the regions given in Table I.
In this table $M_{B_1^*}$ and $M_{B_2^*}$ are treated as free parameters and the restrictions
$M_{B_0^*}=M_{B_1^*}-12$MeV and  $M_{B_1}=M_{B_2^*}-12$MeV are set up.
Also $\Gamma_{B_1^*}$ and   $\Gamma_{B_2^*}$ are free parameters and 
the restrictions  $\Gamma_{B_0^*}=\Gamma_{B_1^*}$ and  $\Gamma_{B_1}=\Gamma_{B_2^*}$
are set up.
It is also taken into account the fact that the effective mass of relevant resonances in the background
processes is shift down by $\Delta M_\gamma = m_{B^*}-m_B=46$MeV.
The upper and lower limits of the regions on the free parameters
$m_{B_1^*}$, $m_{B_2^*}$, $\Gamma_{B_1^*}$ and $\Gamma_{B_2^*}$ are determined
in reference to their original values of L3 also shown in Table I.
\begin{table}[h]
\begin{center}
\caption{Resiticted regions of masses and widths (in MeV)}
\begin{tabular}{clc|clc}
\hline
 & & Direct Res. Process & & & Background Res. Process \\
\hline
$B_0^\chi$ & $m$ & 5530--5570 & $B_1^*$ & $m$ & 5590--5640 \\
         & $\Gamma$ & 10--40 &   &   & (L3; $5670\pm 10\pm 13)$ \\
$B_0^*$ & $m$ & $m(B_1^*)$-12+46 &   &    $\Gamma$ & 20--100 \\
         &   $\Gamma$ &  $\Gamma (B_1^*)$ &    &   & (L3; $70\pm 21\pm 25)$ \\
\hline
$B_2^*$ & $m$ & 5730--5780 & $B_1$ & $m$ & $m(B_2^*)-12-46$ \\
     &    & (L3; $5768\pm 5\pm 6)$ &   &   $\Gamma$ &  $\Gamma (B_2^*)$ \\
    & $\Gamma$  &  0--60  &   $B_2^*$ & $m$ & $m(B_2^*)-46$ \\
    &   & (L3; $24\pm 19\pm 24)$ &  &    $\Gamma$ &  $\Gamma (B_2^*)$ \\ 
\hline
\end{tabular}
\end{center}
\end{table}\\
iii)  Energy resolution of experiments; \ \ \ \ its effects are taken into account through the relation
\begin{eqnarray}
\Gamma_{R_i} &=& \sqrt{\Gamma_{i,{\rm orig.}}^2+\Gamma_{i,{\rm resol.}}^2}\ ,\ \ \label{eq11}
\end{eqnarray}
where $\Gamma_{i,{\rm resol.}}$ is the energy resolution $\sigma$ of L3, $\Gamma_{R_i}$
is the value for our formula, and $\Gamma_{i,{\rm orig.}}$ is the true width of the resonance $R_i$.\\
iv) Production couplings $r_i$; \ \ \ \ following L3, the production ratios among resonances
are assumed to be proportional to the $(2J_i+1)$ ($J_i$ being spin of $R_i$).
Then $r_i$ are obtained from $\Gamma_i$ through the relation
\begin{eqnarray}
r_i^2\Gamma_i & \propto & (2J_i+1)\ .\ \ \label{eq12} 
\end{eqnarray}

\section{Results of Analysis}

The results of our analysis in both cases, with $B_0^\chi$ and without $B_0^\chi$,
are shown in Fig.~1. The fitted curves in comparison with the data are given, respectively,
as follows:\ \ \ 
In the case with  $B_0^\chi [$ without $B_0^\chi ]$ the results on exclusive L3 spectra and
ALEPH ones are shown in Fig. (a), (c) and (e) $[$ (b), (d) and (f) $]$ .

\begin{figure}
\epsfysize=20. cm
  \centerline{\epsffile{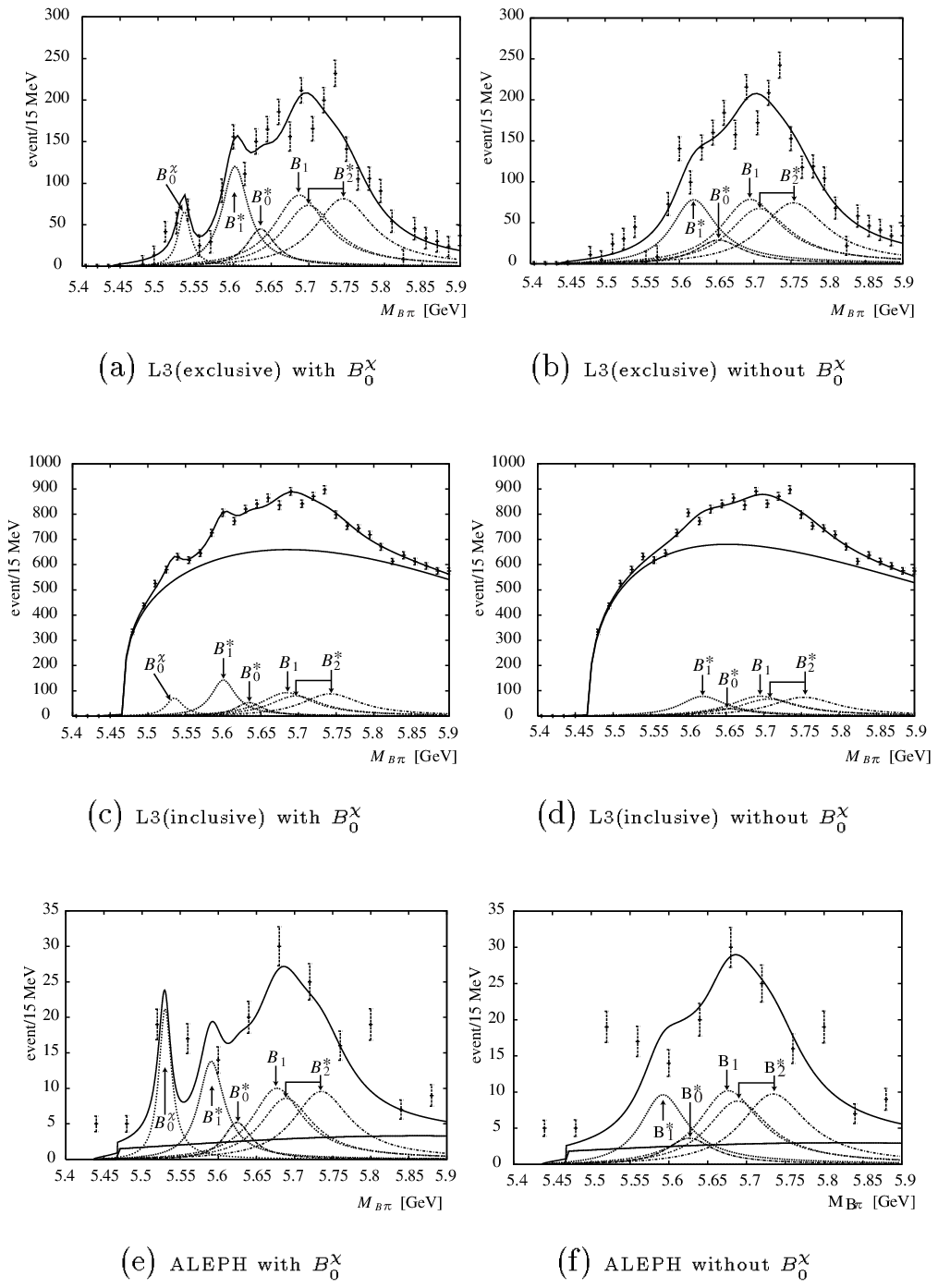}}
\caption{Fitted curves in the case with (without) $B_0^\chi$ and experimental data
of L3(exclusive), L3(inclusive) and ALEPH are shown, respectively, in Fig. 
(a), (c) and (e) ( (b), (d) and (f) ). The contributions from respective resonances are 
also shown.}
\end{figure}
The values of mass and width of direct resonances, $B_0^\chi$, $B_0^*$ and $B_2^*$,
obtained through our analysis of (a) L3(exclusive) are given in Table II.
\begin{table}[h]
\begin{center}
\caption{Fitted values of mass and width (in MeV)}
\begin{tabular}{cccc}
\hline
   & $B_0^\chi$ & $B_0^*$ & $B_2^*$ \\
\hline
$m$          &  5534 & 5635 & 5743 \\
$\Gamma$ & 19.7 & 40.3  & 73.1 \\
\hline
\end{tabular}
\end{center}
\end{table}
\ \ In Table III the values of $\chi^2$ and of 
$\tilde\chi^2(\equiv \chi^2/{\rm No.\ of\ data\ points - No.\ of\ param.})$ 
of the analysis of L3(exclusive) are given in both the cases with and without $B_0^\chi$.
\begin{table}[h]
\begin{center}
\caption{Values of $\chi^2$ and $\tilde\chi^2$}
\begin{tabular}{l|c|c}
\hline
   & with $B_0^\chi$ & without $B_0^\chi$  \\
\hline
$\chi^2$ & 18.5 &  22.4 \\
$\tilde\chi^2$
     & $\frac{\displaystyle 18.5}{\displaystyle 34-9}=0.74$  & $\frac{\displaystyle 22.4}{\displaystyle 34-5}=0.77$ \\
\hline
\end{tabular}
\end{center}
\end{table}
Through the inspection of fitted curves in Fig. 1, we may conclude that our results give some evidence
for existence of $B_0^\chi$, although there is no significant improvement on the reduced $\chi^2$.

\section{Concluding Remarks}

\hspace*{-0.8cm} ({\it Summary of this work})\ \ \ \ 
We have made reanalysis of the $(B\pi )$ mass spectra obtained through the process Eq.~(\ref{eq1}),
by applying the formulas (\ref{eq10}) and (\ref{eq101}).
As the intermediate direct resonance we have taken into account the $B_0^\chi$, a new chiral scalar meson
outside of the conventional level-classification scheme. As a result we have obtained a possible evidence
of $B_0^\chi$ with $m=5530$ MeV and $\Gamma = 19.7$ MeV. The obtained values of reduced $\chi$ square,
$\tilde\chi^2=18.5/(34-9)=0.74$ for the case with $B_0^\chi$ and $22.4/(34-5)=0.77$ for the case without
$B_0^\chi$, respectively. However, the statistical accuracy of the relevant data is very poor,
and the more accurate data are required in order to get a definite conclusion.

\hspace*{-0.8cm} ({\it Universal property of chiral particles}) \ \ \ \ 
In the previous works the universal relations\cite{hl,rfbardeen,rfebert} of chiral particles 
through the $D$- and $B$-meson systems
are predicted from the chiral symmetry on the light-quark and the heavy quark symmetry:
The mass splittings between the respective chiral partners are equal, and the decay widths for the 
one-pion emission of the chiral particles, scalar and axial vector mesons are also equal.

We can check some of these universal relations experimentally, using the results of the analyses 
given in this letter and the previous one\cite{D1chi}, as follows:
\begin{eqnarray}
\Delta m_D (\equiv m(D_1^\chi )-m(D^*)) &=& \Delta m_B (\equiv m(B_0^\chi )-m(B))  \nonumber\\
302(2312-2010)\ {\rm MeV}\ \ \ \    & & 238(=5535-5297)\ {\rm MeV.}\ \ \label{eqC1}\\
     & & \nonumber\\ 
\Gamma (D_1^\chi \rightarrow D^* \pi ) &=& \Gamma (B_0^\chi \rightarrow B\pi )  \nonumber\\
23\ {\rm MeV}\ \ \ \ \ \ \ \    & & 20\ {\rm MeV.}\ \ \label{eqC2}
\end{eqnarray}
From the above we may conclude that the theoretical predictions on the universal properties of chiral
particles are consistent with the present experiment.\\

The authors should like to express their gratitude to Professors, K.~Takamatsu and T.~Tsuru,
for their useful discussions and encouragement.

\end{document}